\documentclass[conference]{IEEEtran}
\IEEEoverridecommandlockouts
\usepackage{cite}
\usepackage{amsmath,amssymb,amsfonts}
\usepackage{graphicx}
\usepackage{textcomp}
\usepackage{xcolor}
\usepackage{enumitem}
\setlist{leftmargin=3mm}

\def\BibTeX{{\rm B\kern-.05em{\sc i\kern-.025em b}\kern-.08em
    T\kern-.1667em\lower.7ex\hbox{E}\kern-.125emX}}
\begin{document}

\title{Using POMDP-based Approach to Address Uncertainty-Aware Adaptation for Self-Protecting Software}

\author{
    \IEEEauthorblockN{Ryan Liu,  Ladan Tahvildari}
    \IEEEauthorblockA{
        \textit{Dept. of Electrical and Computer Engineering} \\
        \textit{University of Waterloo, Canada}\\
        \{ryan.zheng.he.liu, ladan.tahvildari\}@uwaterloo.ca
    }
}

\maketitle

\begin{abstract}
The threats posed by evolving cyberattacks have led to increased research related to software systems that can self-protect. One topic in this domain is Moving Target Defense (MTD), which changes software characteristics in the protected system to make it harder for attackers to exploit vulnerabilities. However, MTD implementation and deployment are often impacted by run-time uncertainties, and existing MTD decision-making solutions have neglected uncertainty in model parameters and lack self-adaptation. This paper aims to address this gap by proposing an approach for an uncertainty-aware and self-adaptive MTD decision engine based on Partially Observable Markov Decision Process and Bayesian Learning techniques. The proposed approach considers uncertainty in both state and model parameters; thus, it has the potential to better capture environmental variability and improve defense strategies. A preliminary study is presented to highlight the potential effectiveness and challenges of the proposed approach.
\end{abstract}


\section{Introduction}
As cyberattacks become more sophisticated and widespread, the need for effective security measures to defend against these attacks becomes increasingly important. Self-Protecting Software (SPS)~\cite{yuanSPSSurvey} is an important topic of research in the field of cybersecurity and software engineering, as it has the potential to significantly enhance the security and resilience of software systems against a wide range of threats and vulnerabilities. One approach to realizing SPS is Moving Target Defense (MTD)~\cite{yuanSPSSurvey,choSPSSurvey}, which involves continuously altering the characteristics of a system in order to make it more difficult for an attacker to exploit vulnerabilities or predict the system's behaviour. However, implementing MTDs can be challenging, as it requires the ability to continuously adapt the defense to the changing circumstances of the attack, and do so in a cost-effective manner that does not disrupt normal system operation. Moreover, the deployment of MTD techniques is often subject to run-time uncertainties that can affect their effectiveness. These uncertainties may include (i) incomplete knowledge of the true state of the environment, (ii) changes in the effectiveness of security countermeasures in the environment (e.g. zero-day attacks), and (iii) variations in the number of distinct attacker groups targeting a given system.

Existing model-based solutions for MTD decision-making have often overlooked the impact of model parameter uncertainty and lack self-adaptation in the face of these run-time uncertainties. Therefore, the main goal of the proposed research is to address this deficiency and to improve the uncertainty-awareness and self-adaptation of MTD decision-making. To achieve this goal, we propose solutions based on Partially Observable Markov Decision Process (POMDP)~\cite{Spaan2012, KAELBLING98, POMCP, BAPOMDPlong} to quantify decision-making uncertainties and to develop an uncertainty-aware MTD decision engine. The resulting decision engine will be able to adapt to dynamic environments with various unknowns. By taking into account the uncertainty in the model parameters, the proposed approach aims to better (i) capture the inherent variability and complexity of the operating environment, and (ii) better coordinate reliable and effective defense strategies. 

\section{Research Motivation} \label{sec:motivation}

\subsection{Moving Target Defense}
\begin{figure}
    \vspace{-5mm}
    \centering
    \includegraphics[width=0.35\textwidth]{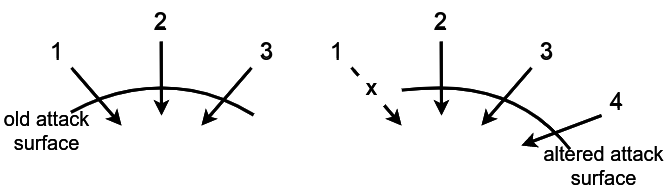}
    \caption[Graphical illustration of MTD technique]{Graphical illustration of MTD technique. (left) there are 3 attack paths for attackers to breach the system. (right) after shifting the attack surface of the system, attack path 1 no longer works. However, the shift may enable new attack paths (i.e. attack path 4)}
    \label{fig:attacksurface}
    \vspace{-5mm}
\end{figure}

The basic axiom of MTD techniques is that it is impossible for defenders to provide complete and perfect security for a given software system~\cite{choSPSSurvey,TORQUATO2020101742}. Therefore, MTD techniques can be defined as cybersecurity techniques that continuously alter the protected system's attack surfaces and configurations in order to increase the effort required to exploit the system's vulnerabilities~\cite{choSPSSurvey,Ward2018SurveyOC}. Here, the term attack surface is defined as the set of ways in which an attacker can enter the system and potentially cause damage. A graphical example of this idea is shown in Figure~\ref{fig:attacksurface}. Altering the attack surface of the protected system can be achieved by introducing heterogeneity, dynamicity, and non-deterministic behaviour~\cite{yuanSPSSurvey}. Some examples of this include shuffling network addresses, rotating encryption keys, or randomly modifying system configurations. 

MTD can be used in a variety of contexts, such as individual devices and networked systems. However, there are also some challenges to using MTD. One of the main challenges is that implementing and incorporating MTD can be complex and time-consuming, as it requires careful planning to ensure that the protected system remains functional and secure while it is continuously being altered. It is important to carefully consider the potential benefits and drawbacks of MTD before deploying it in the operating environment. An example of a potential drawback in MTDs is the addition of new threats when the attack surface is changed (as shown in Fig~\ref{fig:attacksurface}.) Examples of other overheads when deploying MTDs include increased memory usage, decreased throughput, operational costs, etc.~\cite{choSPSSurvey,Ward2018SurveyOC}.


\subsection{Partially Observable Markov Decision Process}

POMDP is a mathematical framework used to model sequential decision-making problems in which an agent must make decisions based on incomplete information about the state of the environment~\cite{Spaan2012,KAELBLING98}. A POMDP problem is defined by a set of states, actions, observations, a transition function, an observation function, and a reward function~\cite{Spaan2012,KAELBLING98}. In solving a POMDP problem, the objective is to learn an optimal policy, which is a mapping from states to actions, that will maximize the expected cumulative reward over time. 

A key challenge in solving POMDPs is dealing with state uncertainty due to incomplete knowledge of the true state of the system. To address this challenge, a belief state, which is a probability distribution over states of the environment, is used to summarize the agent's previous experience~\cite{Spaan2012,KAELBLING98}. Techniques such as filtering and prediction allow the agent to use its belief state to infer the true state of the environment. Overall, there are numerous algorithms that can be used to solve POMDPs, such as exact value iteration, point-based value iteration, and Monte-Carlo (MC) methods~\cite{Spaan2012,KAELBLING98,DESPOT,POMCP,BAPOMCP,FBAPOMDP,Perseus}. These algorithms use different approaches to compute the values (i.e. expected cumulative rewards) of states and actions to find the optimal policy. 


Another challenge related to solving POMDP problems is the issue of model parameter uncertainty (i.e., uncertainty in the model of the system itself). This issue mainly affects model-based POMDP approaches, which rely on a model of the environment to find an optimal policy. Possessing full knowledge about the POMDP model is usually a strong assumption in practice~\cite{BAPOMDPlong}. If decisions are made based on a model that does not perfectly fit with the real problem, the agent may risk executing low return actions~\cite{BAPOMDPlong}. In the context of Reinforcement Learning (RL), this challenge is summarized as the problem of exploration-exploitation (i.e. whether to \textit{explore} the environment and learn more about it or to \textit{exploit} the current knowledge about the environment to maximize the objective). 

\subsection{Research Gaps}
While there have been many approaches proposed for improving MTD deployment, most have focused on addressing state uncertainty~\cite{EghtesadLong,Prakash2015,mcabee_tummala_mceachen_2021} and have overlooked the impact of model parameter uncertainty. The main research gaps in the existing literature are:
\begin{itemize}
    \item \textbf{Lack of consideration of model parameter uncertainty during decision-making process}
    \item \textbf{More self-adaptive MTD mechanisms can be developed to adapt to changing attacker behaviour and defense efficiency}
    \item \textbf{Lack of scalability in modelling multiple attackers and simultaneous attacks across various system components}
\end{itemize}

The proposed approach is presented in the next section based on these motivations.
\section{Proposed Approach} \label{sec:approach}
\begin{figure}[!b]
\vspace{-5mm}
    \includegraphics[width=.45\textwidth]{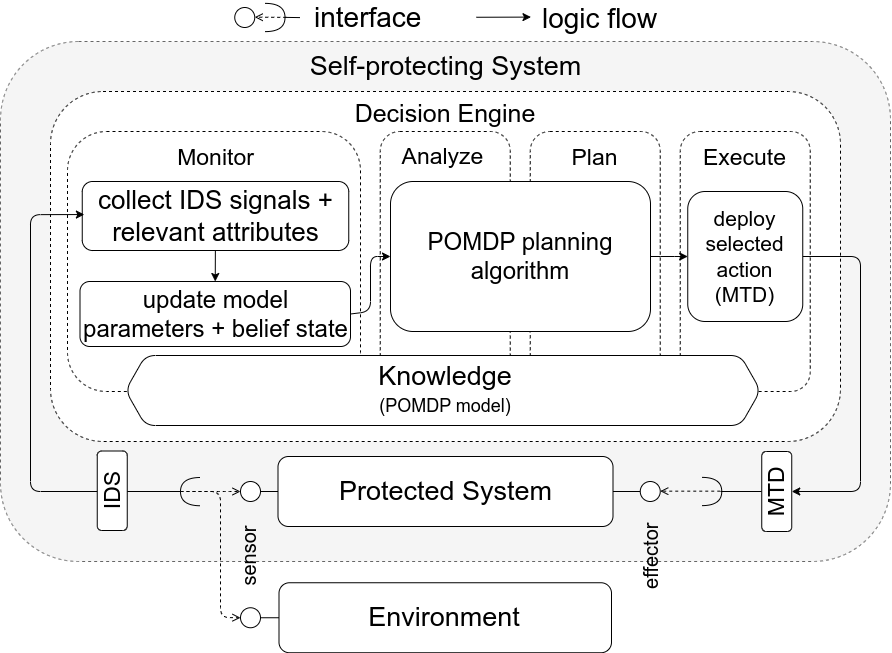}
    \centering
    \caption{Overview of POMDP-based MTD decision engine}
    \label{fig:adaptationLoop}
\end{figure}

The main novelty of the proposed approach is the consideration of model parameter uncertainty and enhanced self-adaptation during the planning process. By taking into account both types of uncertainties, this approach will be able to better respond to the complexity and unpredictability of real-world scenarios. Figure~\ref{fig:adaptationLoop} presents an overview of the proposed decision engine.

\vspace{-1mm}

\subsection{Assumptions}
In designing the proposed approach, the following general assumptions are made regarding the capabilities of the defender and attacker(s). 

\begin{itemize}[leftmargin=3mm]
    \item The protected system is initially under defender control.
    \item The true security status of the protected system is not known by the defender at all times.
    \item The effectiveness of MTDs and intrusion detection systems (IDSs) may vary over time.
    \item The attacks follow a sequential series of phases, which ultimately lead to compromising the protected system.
    \item The attack progression is at least partially observable to the defender.
    \item The number of attackers and attacker activities in the operating environment may vary over time.
\end{itemize}


\subsection{Modelling}
In order to decide on an appropriate series of actions for self-protection, a comprehensive model of the system is required. The model will capture both domain and uncertainty-related knowledge in a compact manner that can be utilized for MTD decision-making.

\begin{figure}
    \includegraphics[width=0.4\textwidth]{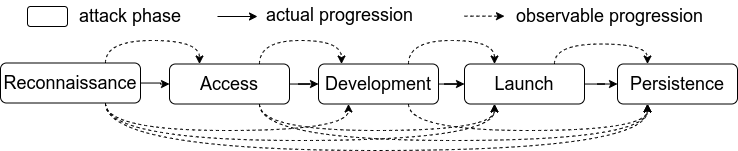}
    \centering
    \caption[An attack kill chain]{An attack kill chain. The solid arrows show the actual progression path between phases as the adversary advances in their attack. The dashed arrows show the observable progression paths between phases (by the defender) due to the partially observable nature of the domain. }
    \label{fig:killChain}
\vspace*{-5mm}
\end{figure}

\subsubsection{Capturing the Domain}
Fundamentally, all MTD techniques are focused on disrupting attack progression~\cite{Ward2018SurveyOC}. Disruption of an attack is accomplished by introducing dynamicity to make the protected system less deterministic and homogeneous~\cite{Ward2018SurveyOC}. In this regard, all MTD techniques can be categorized based on the phase of attack they intend to disrupt along an \textit{attack kill chain}. Various versions of the attack kill chain exist in literature and may be applied based on specific MTD domain requirements~\cite{Ward2018SurveyOC}. The proposed approach will rely on a kill chain similar to Figure~\ref{fig:killChain} as the basis for domain modelling during MTD decision-making.



Building on the kill chain as the backbone, analyzing the high-level relationships between domain concepts can serve as a starting point for further domain modelling in the POMDP. The important relationships considered here are:

\begin{itemize}[leftmargin=2.75mm]
    \item \textbf{Quality Goals and Attacker Goals}: 
    For instance, Quality Goals for the protected system could include maintaining availability, throughput, or control-flow integrity, while Attacker Goals could include denial-of-service, data leakage, and identity theft. Quality Goals and Attacker Goals can be used here to derive the reward function in the POMDP. These goals can also assist with the selection of metrics that should be monitored, how the monitored metrics should map to the phases of the attack kill chain, and which MTD techniques should be incorporated as actions available to the agent at run-time. 
    
    \item \textbf{MTDs, IDSs, and Attacks}: The set of applicable MTDs, IDSs, and potential attacks will vary based on the design and development of the protected system. For example, the choice of the programming language used to develop the protected system has a significant impact on the set of MTDs, IDSs, and attacks that should be considered. The relationship between these components in the model also determines the various probabilistic dynamics of the POMDP model. Domain knowledge regarding these concepts based on historical data can serve to configure the initial model parameters in the POMDP. The set of MTDs also dictates the size of the action space in the POMDP model.
    
    \item \textbf{Security Goals and Non-Security Goals}: Quality goals can be further divided into security and non-security goals, which may often be in conflict with one another. For example, an N-variant system can be deployed to run multiple variants of the same application simultaneously and monitor the variants for divergence. The security goal motivating this MTD technique is to mitigate code injection and control injection attacks~\cite{Ward2018SurveyOC}. However, deploying N-variant systems requires a trade-off between system throughput and resource utilization overhead. In this case, throughput decreases and resource utilization overhead increases linearly as the number of variants goes up. 
    
\end{itemize}

\subsubsection{Capturing Uncertainty}
Uncertainties in the decision-making process for MTD techniques can be grouped into two categories: uncertainty regarding the true state of the environment (i.e. state uncertainty) and uncertainty regarding the true dynamics of the environment (i.e. model parameter uncertainty). For both categories, a common source of uncertainty in the context of MTD decision-making is the fact that the defender cannot always discern malicious activities from benign activities in the environment. In consequence, the defender is unable to track the progress of attacker activities with full confidence, which leads to state uncertainty. The defender also cannot know the effectiveness of IDS  mechanisms and deployed MTD techniques against all attacks (e.g. zero-day attacks) with complete certainty; hence, leading to model parameter uncertainty. Figure~\ref{fig:killChain} also illustrates the consequence of state uncertainty with respect to the transitions in attack progression along the attack kill chain. 

In the proposed approach, \textit{state uncertainty} is addressed by introducing partial observability to a discrete MDP to derive a discrete POMDP. As states cannot be derived from observations and direct mappings from observation to actions are insufficient for optimal decision-making, an alternative Markovian signal for choosing actions is needed~\cite{Spaan2012,KAELBLING98}. To overcome this issue, belief states can be used. A belief state is a probability distribution over the state space of the POMDP. It provides a sufficient basis for acting optimally under state uncertainty~\cite{Spaan2012,KAELBLING98}. 

To address \textit{model parameter uncertainty}, the proposed approach intends on exploring the application of Bayesian Learning methods. In particular, concepts from Bayes-Adaptive POMDP (BA-POMDP)~\cite{BAPOMDPlong} can be applied to address model parameter uncertainty. The BA-POMDP framework introduces Dirichlet distribution experience counts into the POMDP state space in order to represent uncertainty regarding the transition function $T(s,a,\cdot)$ and observation function $O(s',a,\cdot)$. Hence, the BA-POMDP model tracks the experience counts $\phi_{ss'}^a  \: \forall s'$, which denotes the number of times transition to state $s'$ after performing action $a$ in state $s$, and $\psi_{s'o}^a \: \forall o$, which denotes the number of times observation $o$ was made in state $s'$ after performing action $a$~\cite{BAPOMDPlong}. 

If we let $\phi$ and $\psi$ be the vector of all transition counts and all observation counts respectively, it can be shown that the expected transition probability for $T(s,a,s')$ and the expected probability for observation $O(s',a,o)$ are:

\begin{equation}
\label{eq:bapomdpPhi}
T_{\phi}(s,a,s') = \frac{\phi_{ss'}^a}{\sum_{s'' \in S}\phi_{ss''}^{a}}
\end{equation}
\begin{equation}
\label{eq:bapomdpPsi}
T_{\psi}(s',a,o) = \frac{\psi_{s'o}^a}{\sum_{o' \in \Omega}\psi_{s'o'}^{a}}
\end{equation}

Building from this, the BA-POMDP framework incorporates the $\phi$ and $\psi$ vectors into the state space to derive a new model. The new set of states (hyperstates in the format $(s,\phi,\psi)$) is defined as $S' = S \times \mathcal{T} \times \mathcal{O}$ where $\mathcal{T}~=~\{ \phi \in \mathbb{N}^{|S|^2|A|} \: | \: \forall (s,a) \in S \times A, \sum_{s' \in S}\phi_{ss'}^{a} > 0 \}$ and $\mathcal{O}~=~\{ \psi \in \mathbb{N}^{|S||A||\Omega|} \: | \: \forall (s',a) \in S \times A, \sum_{o \in \Omega} \psi_{s'o}^{a} > 0 \}$. The action space remains the same as in the POMDP. The state transition and observation functions become functions of the hyperstate, while the reward function remains a function of state-action pairs.




\subsection{Planning}
When planning under uncertainty, utilizing belief states allows the agent (i.e. decision engine) to take into account its own degree of uncertainty about state identity. For example, the belief state of an agent that is \textit{very confident} about the current state of the environment will have few states with high probability and many states with low or zero probability. 

Unfortunately, solving POMDP planning problems exactly with belief states is usually intractable outside of very small domains. However, various approximate algorithms have been proposed to tackle problems with large domains. In particular, online planning via MC tree search is a proven approach that the proposed approach intends to incorporate. The main advantage of MC-based planning compared to other online planning approaches lies in its method for policy evaluation (i.e. value function updating)~\cite{Sutton1998,POMCP}. MC-based planning uses MC sampling to update the value function by sampling start states from the belief state and sampling histories (i.e. observation-action sequences) from a black box simulator (instead of relying on explicit probability distributions)~\cite{POMCP}. By doing so, MC planning can avoid full-width computation in the search tree that non-MC planning algorithms must carry out. The amount of MC simulations can also be reduced by incorporating domain knowledge into the search algorithm as a set of preferred actions. In the context of MTD decision-making under run-time uncertainty, using the augmented hyperstates in BA-POMDP provides further motivation for applying MC planning techniques to improve planning scalability. 

Another computationally expensive task during POMDP planning is the belief state update procedure based on the observation received after performing an action. This update can be achieved by Bayes' theorem, which requires iteration over the state space. As the state space of the POMDP increases, even a single belief state update may become computationally infeasible. To overcome this issue, MC sampling can also be applied to the belief state update procedure. MC belief state update utilizes a particle filter with $K$ particles to approximate the belief state. Each particle represents a sample state $B_{t}^{i} \in S, \: 1 \leq i \leq K$. The approximated belief state is defined as the sum of all particles.

\begin{equation}
    \hat{b}(s) = \frac{1}{K}\sum_{i=1}^{K}\delta_{sB_{t}^{i}}
\end{equation}

Here, $\delta_{sB_{t}^{i}}$ is the Kronecker delta function. $K$ particles are initially sampled from the initial state distribution. Given a real action and observation $o_t$, MC belief update works by sampling a state $s$ from the current belief state (i.e. random selection of a particle from $B_t$). The particle is then passed into the black box simulator to generate a successor state $s'$ and observation $o'$. If the sample observation and real observations match (i.e. $o_t = o'$), the successor state $s'$ is added as a new particle to $B_{t+1}$. This overall procedure is repeated until $K$ particles have been added.

\section{Research Questions}
The main research questions we intend to explore in this preliminary study are:

\textbf{RQ1} \textit{How effective is decision-making using MC-based POMDP planning?} --- The effectiveness of MTD decision-making can be judged from various aspects. We aim to measure the effectiveness of MC-based POMDP planning with respect to the satisfaction of security and non-security goals in the context of the problem domain.

\textbf{RQ2} \textit{What is the impact of neglecting model parameter uncertainty on MTD decision-making?} --- For this study, we only captured state uncertainty and did not incorporate model parameter uncertainty concepts explicitly. However, the focus here is to lay out the foundations for the proposed approach by evaluating decision-making when model parameter uncertainty is \textit{neglected}.

\section{Experimental Evaluation}

\subsection{Experiment Setup}
As a proof of concept, a precursor to the proposed approach, utilizing Partially Observable Monte-Carlo Planning (POMCP)~\cite{POMCP}, was developed and evaluated. The experiments simulate cryptojacking attacks, which involve attackers using compromised machines to mine cryptocurrency without the victim’s consent. With the popularization of cloud computing and containerization, recent cryptojacking attacks, such as Hildegard~\cite{hildegard}, have started to focus on cloud computing resources. Therefore, the experiments mimic protecting nodes in a Kubernetes-based cluster from cryptojacking. Here, we make the following assumptions regarding the capabilities of the attacker and defender:

\begin{itemize}
    \item The attacker can simultaneously attack multiple nodes in cluster
    \item The attack progress only moves in the forward direction along the attack kill chain
    \item The defender can counter attacks by reimaging nodes (restore to pristine state after being offline for a certain amount of time)
    \item The defender can fully observe the availability of each node (i.e. online or offline)
    \item The defender can partially observe the progress of attacks (at a constant \textit{observation rates})
    \item The defender can choose to not reimage any nodes
    \item The reimaging process is deterministic (always follows Up → Shutdown → Bootup → Up)
    \item The nodes take constant time to reimage
    \item Observing attack progress is only possible when attacks advance to the next phase of the attack kill chain  
\end{itemize}

Table~\ref{tab:hildegardDomainInfo} presents the information related to the evaluation domain. In all experiments, the number of nodes in the cluster was limited to 3 (i.e. $N=3$). Based on this information, a POMDP problem was formulated and used during the experiments. In order to provide a basis for comparison, a naive rule-based planner was also implemented. Using the definitions in Table~\ref{tab:hildegardDomainInfo}, the rule-based planner follows a simple configurable rule specified by a parameter $rb \in P$, which indicates when the defender should act to reimage a node given the observed progress $op$ on that node is greater than or equal to $rb$ (e.g. when $rb=1$, the rule-based defender will always reimage a node $i\in[1,N]$ whenever $op_i \geq 1$ (TargetScan)).

\subsection{Obtained Results}

\begin{figure}[!b]
    \vspace*{-5mm}
    \includegraphics[width=0.5\textwidth]{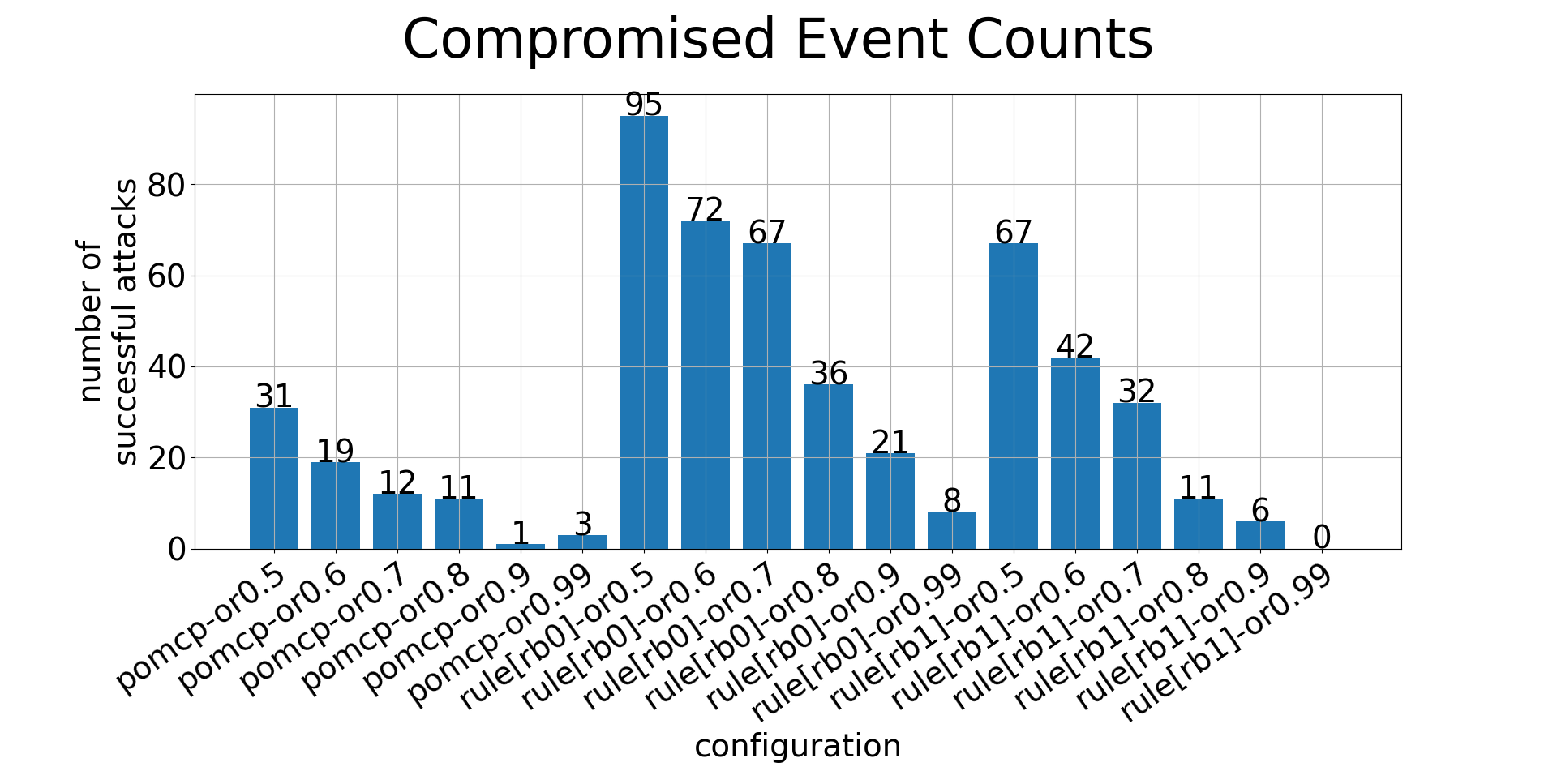}
    \centering
    \caption[Comparison of compromise event counts]{Comparison of compromise event counts}
    \label{fig:compromiseCompare}
    \vspace*{-5mm}
\end{figure}

\subsubsection{\textbf{RQ1} How effective is decision-making using MC-based POMDP planning?} 
\begin{figure}
    \vspace{-2mm}
    \includegraphics[width=0.5\textwidth]{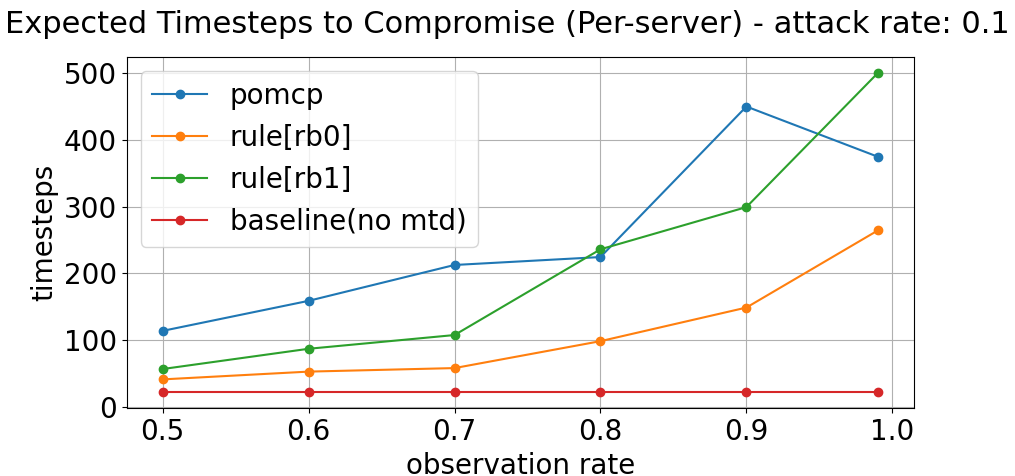}
    \caption[Comparison of expected time steps until compromise]{Comparison of expected time steps until compromise}
    \label{fig:expectedTimestepsUntilCompare}
    \vspace{-2mm}
\end{figure}

From the perspective of the security goal, we can observe from Figure~\ref{fig:compromiseCompare} that the total number of compromise events (i.e. the number of times attacker successfully deployed their cryptomining process) for the POMCP planner is generally lower than the corresponding rule-based planners across the range of observation rates. In addition, Figure~\ref{fig:expectedTimestepsUntilCompare} also illustrates that the POMCP planner is able to extend the average amount of timesteps (i.e. effort) required by the attacker to successfully compromise a node in the cluster.

\begin{figure}
    \vspace*{-5mm}
    \includegraphics[width=0.45\textwidth]{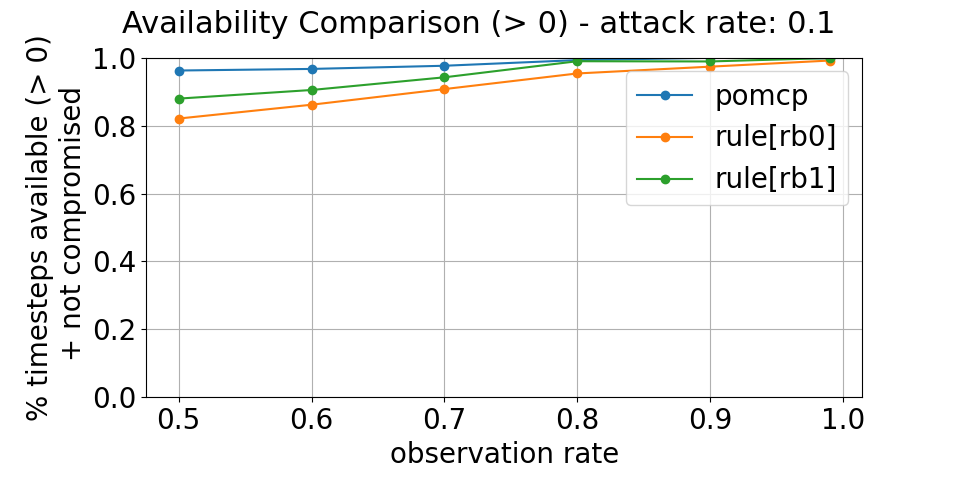}
    \centering
    \caption{Comparison of cluster availability with at least 1 node (out of 3)}
    \label{fig:availCompare1}
    \vspace{-2mm}
\end{figure}

\begin{figure}
\vspace{-2mm}
    \includegraphics[width=0.45\textwidth]{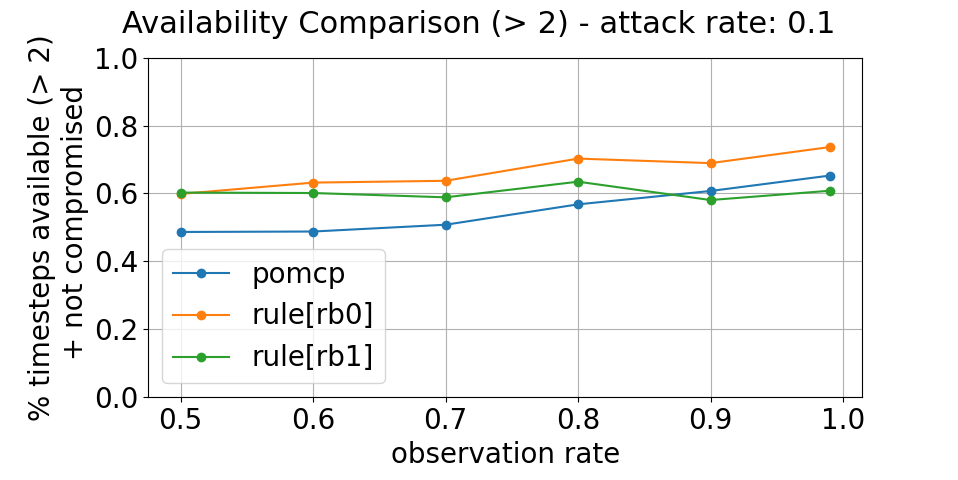}
    \centering
    \caption{Comparison of cluster availability with 3 nodes (out of 3)}
    \label{fig:availCompare2}
\vspace{-4mm}
\end{figure}

From the perspective of the non-security goal, we can observe the percentage of time there was \textit{at least one node} \textit{available} and \textit{uncompromised} in the cluster in Figure~\ref{fig:availCompare1}. Here, the ~POMCP planner proved to be sufficient in meeting the objective of maximizing availability for benign users relative to the other planners. The only evaluation dimension in which the POMCP solution consistently trailed the rule-based planners is the percentage of time when all nodes were available and uncompromised during the experiments in Figure~\ref{fig:availCompare2}. Regarding this, one can reasonably argue the sacrifices made here are justified based on the enhanced protection offered by the POMCP planner.

\begin{table}[!t]
\vspace{-2mm}
\caption{Domain information for the experiment (Assuming a cluster with $N$ nodes)}
\label{tab:hildegardDomainInfo} 
\begin{tabular}{p{0.12\textwidth} p{0.34\textwidth}}

\hline
\textbf{Domain \newline Information} & \textbf{Description}\\
\hline

 Relevant Attributes &  For each node $i~\in~[1,N]$, the attributes and observed attributes are $attr_i~=~\langle~p_i,~c_i,~av_i,~o_i~\rangle$ and $oattr_i=\langle~av_i,op_i~\rangle$ where, \\
  (attack progress)  &  $p_i \in P = \{ \text{None = 0}, \text{TargetScan = 1}, \text{Launched = 2} \}$ \\
  (compromised)  &  $c_i \in C = \{ \text{False = 0}, \text{True = 1} \}$ \\
  (availability)  &  $av_i \in AV = \{ \text{Up = 0}, \text{Shutdown = 1}, \text{Bootup = 2} \}$ \\
  (obs. progress)  &  $op_i \in P = \{ \text{None = 0}, \text{TargetScan = 1}, \text{Launched = 2} \}$ \\  \hline
MTD~Countermeasures ($C$) & The only MTD countermeasure considered is to \textit{reimage} a node (i.e. restore a node to a pristine state by taking it offline for a certain amount of time). \\ \hline
IDSs~($M$) & Indicators of compromise for the Hildegard attack include the presence of files named "install\_monerod.bash" and "sGAU.sh"~\cite{hildegard}. IDS techniques to detect network scanning~\cite{NetworkScanDetect,PortScanDetect} are also applicable here. \\ \hline
Quality~Goals ($QG$) & The security goal for the defender is to prevent attackers from using nodes for unwarranted cryptocurrency mining. The non-security goal is to maximize the availability of the cluster. \\ \hline
Attacker~Goals ($AG$) & The goal of the attacker is to launch mining scripts on cluster nodes while remaining hidden from IDS.\\ \hline
Attacks ($AT$) &  The attacks deployed by the attacker include port scanning, defense evasion (e.g. library injection, ELF binary encryption), and cryptojacking. \\ \hline
\end{tabular}
\vspace*{-4mm}
\end{table}

\subsubsection{\textbf{RQ2} What is the impact of neglecting model parameter uncertainty on MTD decision-making?} To evaluate the impact of neglecting model parameter uncertainty on POMCP, experiments were carried out where noise was applied to the attack rate parameter of the model, which represents the speed of attacker progression in the system. 

\begin{figure}
    \vspace*{-1mm}
    \includegraphics[width=.5\textwidth]{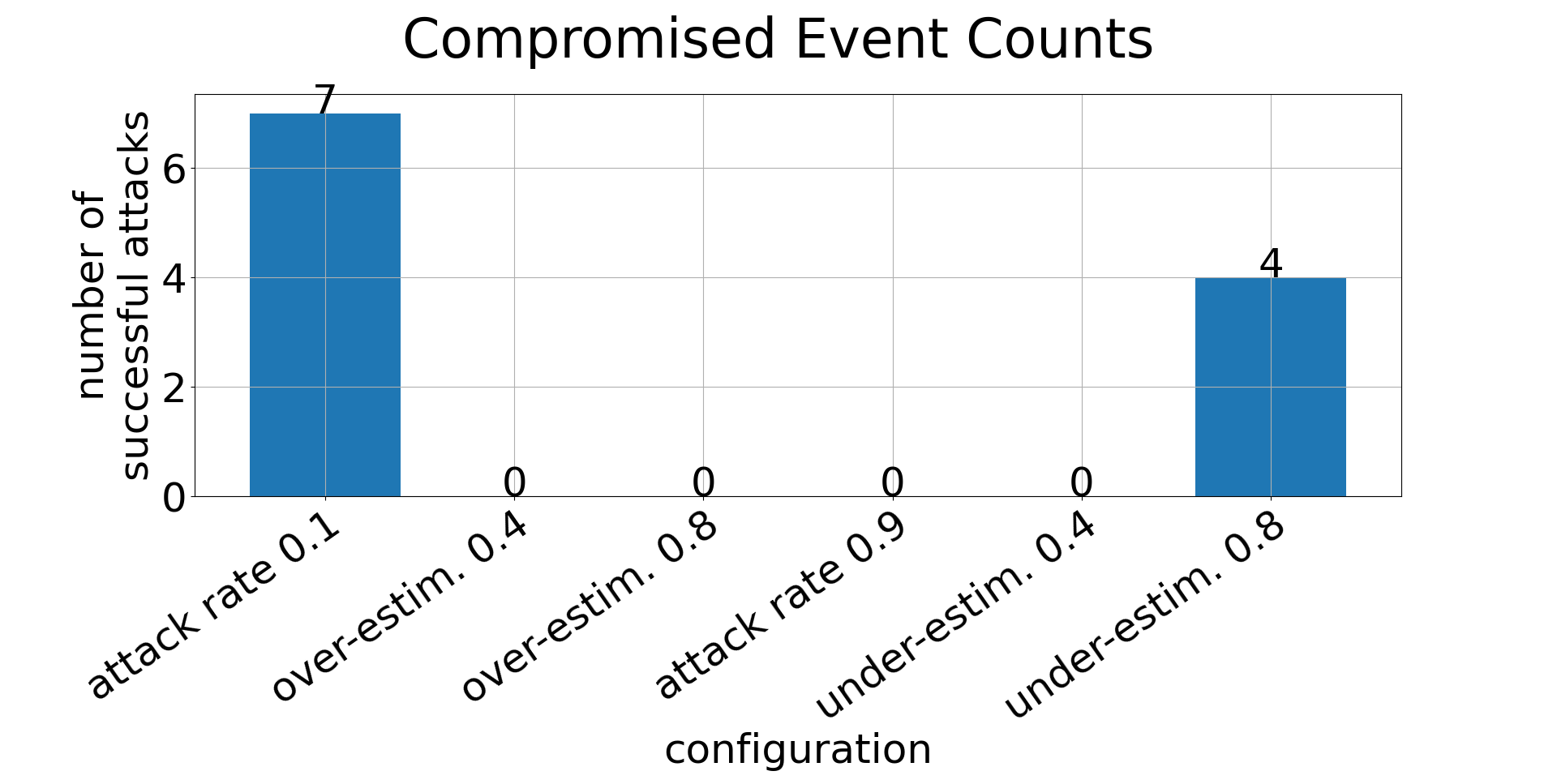}
    \centering
    \caption{Compromise event counts under model parameter uncertainty}
    \label{fig:noisycomp}
    \vspace*{-7mm}
\end{figure}

\begin{figure}[!b]
\vspace{-5mm}
    \includegraphics[width=0.5\textwidth]{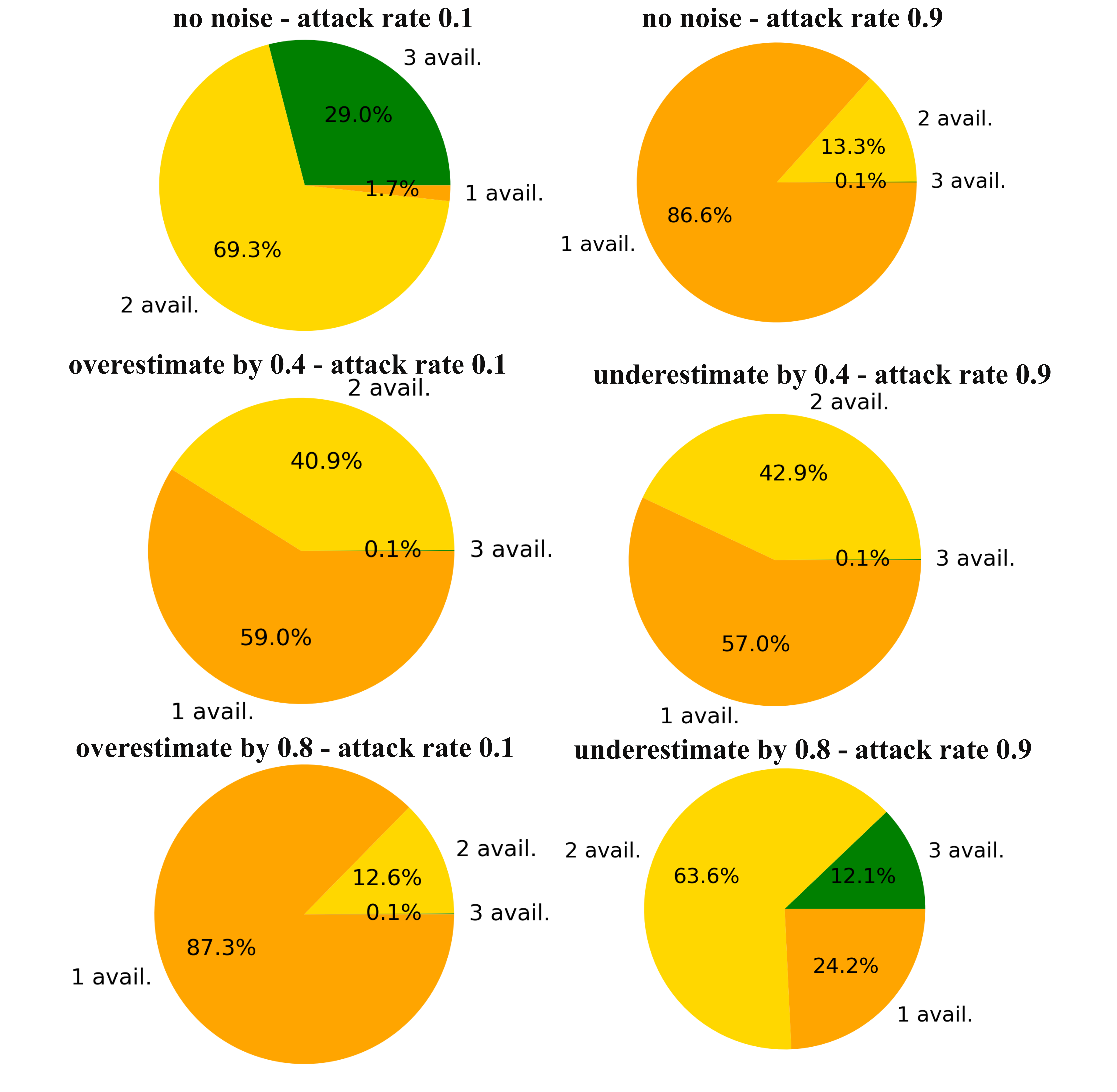}
    \centering
    \caption{Cluster availability distribution comparison under model uncertainty (3 nodes). Pie charts represent the percentage of time nodes are available.}
    \label{fig:noisyavail}
    \vspace*{-5mm}
\end{figure}

The three bar charts on the left in Figure~\ref{fig:noisycomp}, show the result of experiments where the true attack rate was 0.1 but the attack rate used during the POMCP simulations was 0.1, 0.5, and 0.9 respectively. The initial observation here suggests that the noisy models were able to outperform the accurate model. However, upon taking into account the 3 pie charts in the left column of Figure~\ref{fig:noisyavail}, we can notice that the decrease in compromise events for the noisy models can be attributed to the fact that nodes were frequently reimaged (offline) in the cluster. Intuitively, this aligns with the fact that positive noise injected in these experiments triggered the planner to \textit{overestimated} the capabilities (i.e. speed) of the adversary. As a result, while this means the attacker can no longer compromise the node, it also means benign users will suffer a drastically reduced quality of service. 

On the other hand, the right three bar charts in Figure~\ref{fig:noisycomp} are related to the outcome when the defender \textit{underestimates} the capabilities of its adversaries. In this case, planning based on the model with a noise value of -0.8 consistently resulted in particle deprivation during experiments. Overall, the results in compromise event count and cluster availability trend in the opposite direction of the overestimating scenarios.

\subsection{Limitations}
In terms of limitations, more effort is required to compare the proposed approach with more advanced, state-of-the-art techniques found in the literature to identify areas of strengths and weaknesses. Furthermore, the results presented in this paper are not at the scale of real-world scenarios (only 3 nodes were considered in the experiments). The main challenge here is how to capture relevant information in the environment without state space explosion. For the POMCP-based solution in this preliminary experiment, a consequence of increased state space is the occurrence of particle deprivation during the planning process. The current workaround for this is to increase the number of simulations and particles during experiments. However, this obviously will not scale. A potential direction for future exploration is the integration of Factored POMDPs~\cite{FBAPOMDP,factoredPOMDP}. In addition, as shown from the experimental results presented for RQ2, neglecting model parameter uncertainty has the potential to be detrimental to the satisfaction of security and non-security goals. The next step here is to incorporate Bayesian Learning techniques~\cite{BAPOMDPlong,BAPOMCP} into the planning process and analyze the impact of model parameter uncertainty-awareness in the decision-making outcome.

\section{Related Works}
There exist some interesting approaches for MTD planning in current literature, such as using evolutionary algorithms~\cite{John2014} and renewal reward theory~\cite{Wang2016}. However, a widely popular method for MTD planning is the application of Game Theory (GT). One study applied GT to optimize MTD platform diversity and showed that deterministic strategies leveraging fewer platforms can increase system security~\cite{Carter2014}. Research using Empirical GT analysis to model interactions between defender and attacker has also been explored~\cite{Prakash2015}. Furthermore, Empirical GT has also been applied to analyze the effectiveness of MTD in defending against Distributed Denial of Service attacks~\cite{Wright2016}. Another GT-based approach applies Bayesian Stackelberg games for decision-making that balances between computational performance and data privacy~\cite{Bai2019}. The authors model the interactions between cloud providers and users in a risk-aware manner by considering each party's risk preferences. A similar study also utilized Bayesian Stackelberg games to show that MTDs can be improved when combined with information disclosure via a signalling scheme~\cite{Feng2017}. In addition, a recent study proposed using Dynamic Markov Differential Game Model that considers the interplay between the defender and the attacker in MTD~\cite{Zhang2020}. The model takes into account the dynamics of the system state and the impact of defense decisions on the system's behaviour. Nonetheless, one limitation to many GT-based approaches is the consideration of only a single attacker or a single protected system in their analysis.

Reinforcement Learning (RL) techniques have also been extensively studied for MTD planning. For example, RL algorithms were used for generating adaptive strategies in real-time to defend against Heartbleed attacks~\cite{Zhu2014}. RL has also been applied to derive dynamic configuration strategies for the live migration of Linux containers based on a predator (attacker) vs prey (containers) game~\cite{Azab2016}. Another study utilized RL to develop a cost-effective and adaptive defense against multi-stage attacks~\cite{Hu2020}. Moreover, Multi-objective RL has also been used to optimize the attack surface and configuration diversity of computer systems~\cite{Tozer2015}. Deep RL and multi-agent POMDP modelling has been explored to increase the difficulty of exploiting system vulnerabilities~\cite{EghtesadLong}. A POMDP-based approach using attack kill chains has also been explored in a recent study that aims to determine the optimal trade-off between defense effectiveness and costs~\cite{mcabee_tummala_mceachen_2021}. With exception to ~\cite{Hu2020}, a key aspect missing from the aforementioned RL-based approaches is the consideration of model parameter uncertainty.

\vspace{-3mm}
\section{Conclusion}
In summary, this paper presents a novel approach to MTD decision-making based on POMDP modelling with additional run-time uncertainty-awareness via Bayesian Learning techniques. The primary novelty presented here lies in the consideration of model parameter uncertainty on top of state uncertainty. The initial outcomes from the preliminary experiments provide encouraging evidence regarding the plausibility of this research. Nevertheless, further in-depth investigations and evaluations regarding the proposed approach are required before its full potential can be realized.





\vspace{12pt}

\end{document}